\documentclass[aps,prl,twocolumn,superscriptaddress]{revtex4-2}

\usepackage{graphicx}
\graphicspath{{img/}}
\usepackage{amsmath,amssymb,mathtools}
\usepackage{braket}
\usepackage{booktabs}
\usepackage{iftex}
\ifXeTeX
    \usepackage{xeCJK}
    \DeclareRobustCommand{\chinesename}[1]{#1}
\else
    \usepackage{CJKutf8}
    \DeclareRobustCommand{\chinesename}[1]{\begin{CJK*}{UTF8}{gbsn}#1\end{CJK*}}
\fi
\usepackage{bm}
\usepackage{xcolor}
\usepackage{dsfont}
\usepackage[ruled]{algorithm2e}
\SetKwInOut{Input}{Input}
\SetKwInOut{Output}{Output}
\usepackage{hyperref}

\begin{document}

\title{Combining Schr\"odinger and Heisenberg Evolution to Extend Quantum Many-Body Simulations}

\author{Zhigang Hu (\chinesename{胡志刚})}
\affiliation{Wilczek Quantum Center, Shanghai Institute for Advanced Studies, University of Science and Technology of China, Shanghai 201315, China}
\affiliation{Hefei National Laboratory, Hefei 230088, China}

\author{Igor Ermakov}
\email{ermakov1054@yandex.ru}
\affiliation{Wilczek Quantum Center, Shanghai Institute for Advanced Studies, University of Science and Technology of China, Shanghai 201315, China}
\affiliation{Hefei National Laboratory, Hefei 230088, China}
\affiliation{Department of Mathematical Methods for Quantum Technologies, Steklov Mathematical Institute of Russian Academy of Sciences, 8 Gubkina St., Moscow 119991, Russia.}

\author{Yauheni Talochka}
\affiliation{Meshcheryakov Laboratory of Information Technologies, Joint Institute for Nuclear Research, Joliot-Curie Str. 6, Dubna, 141980, Moscow Reg., Russia}

\author{Biao Wu (\chinesename{吴飙})}
\affiliation{International Center for Quantum Materials, School of Physics, Peking University, Beijing 100871, China}
\affiliation{Wilczek Quantum Center, Shanghai Institute for Advanced Studies, University of Science and Technology of China, Shanghai 201315, China}
\affiliation{Hefei National Laboratory, Hefei 230088, China}

\author{Oleg Lychkovskiy}
\affiliation{Skolkovo Institute of Science and Technology, Bolshoy Boulevard 30, bld. 1, Moscow 121205, Russia.}
\affiliation{Department of Mathematical Methods for Quantum Technologies, Steklov Mathematical Institute of Russian Academy of Sciences, 8 Gubkina St., Moscow 119991, Russia.}

\date{\today}

\begin{abstract}
Classical simulations of quantum many-body dynamics face complementary
limitations: entanglement growth restricts state evolution in the
Schr\"odinger picture, while operator spreading makes Heisenberg-picture
calculations increasingly expensive. Here, we combine both approaches
by evolving quantum states and observables independently and joining
their dynamics through massively parallel tensor-network contractions. We implement this
strategy using tree tensor networks and sparse Pauli propagation,
extending evolution times beyond those reached by either method
separately in our calculations. For a quench of the two-dimensional
transverse-field Ising model at the critical field, we simulate the
magnetization of an $11\times11$ open lattice up to $t=1.20$ by combining
state and operator evolution over intervals $t_S=t_H=0.60$.
We compare the dynamics with iPEPS, numerical linked-cluster expansion,
and X-truncated sparse Pauli dynamics results.
We also combine exact state evolution on small clusters with operator
propagation on a larger lattice, improving agreement with iPEPS relative
to cluster evolution alone.
We further derive error bounds showing how the state can suppress
operator errors and the evolved observable can reduce sensitivity to
state errors.
\end{abstract}

\maketitle

\paragraph{Introduction.}

Accurate simulation of quantum many-body dynamics is one of the central tasks in modern computational physics. Reaching longer simulation times is important for answering fundamental questions of nonequilibrium physics, such as when and under what conditions isolated quantum many-body systems equilibrate \cite{eisert2015quantum,rigol2008thermalization,Banuls_2011_Strong,Han2015,Hu2019PRE,ermakov2025classical}, as well as for establishing reliable classical benchmarks for assessing claims of quantum advantage \cite{kim2023evidence,tindall2024efficient,beguvsic2024fast}. Tensor-network methods in the Schrödinger picture are well established \cite{paeckel2019time,orus2019tensor,tindall2024efficient,krinitsin2025time,pavevsic2026scattering,wu2026real,QTensorNet2026}, but rapid entanglement growth limits accessible simulation times, particularly in two and three dimensions. In parallel, Heisenberg-picture approaches \cite{uskov2024quantum,ermakov2024unified,ermakov2025polynomially,beguvsic2024fast,beguvsic2025real,Loizeau2025,miller2025simulation,Rakovszky_2022_Dissipation-assisted,loizeau2025opening,rudolph2026pauli,ermakov2025symbolic,ermakov2025operator,shirokov2025quench,lychkovskiy2026qcommute,SciPostPhysCodeb.84-r1.0,orfi2026scalable,broers2026scalable,pineda2026pauli,miller2026monoprop} have advanced rapidly, enabling increasingly large-scale simulations of quantum many-body dynamics in regimes where strong entanglement growth poses a major challenge to Schrödinger-picture methods.

\begin{figure}[t]
    \centering
    \includegraphics[width=\columnwidth]{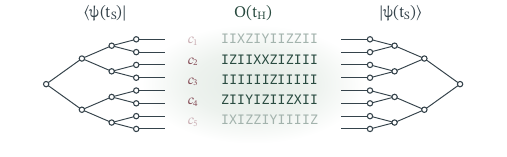}
    \caption{Schematic representation of the hybrid method.
    The trees represent the TTN state evolved to time $t_S$ in the
    Schr\"odinger picture and its conjugate. The central cloud represents
    the operator evolved to time $t_H$ in the Heisenberg picture,
    expressed as a weighted sum of Pauli strings.}
    \label{fig:hybrid-scheme}
\end{figure}

A natural idea is to combine Schrödinger- and Heisenberg-picture evolution to extend accessible simulation times. Suppose the state can be evolved reliably to time $t_S$, $|\Psi(t_S)\rangle=U(t_S)|\Psi_0\rangle$, and the observable to time $t_H$, $O(t_H)=U^\dagger(t_H)OU(t_H)$. The identity
\begin{equation}
    \langle O(t_S+t_H)\rangle
    =\langle\Psi(t_S)|O(t_H)|\Psi(t_S)\rangle
\end{equation}
then suggests that combining the two methods could give access
to the sum of their individually accessible times. The idea is illustrated
in Fig.~\ref{fig:hybrid-scheme}. Achieving this gain in practice, however, is not trivial: evaluating the expectation value introduces an additional computational cost, and errors from both evolutions must be controlled. This idea and strategies for implementing it are discussed in Ref.~\cite{rudolph2026pauli}, which refers to this approach as ``meet-in-the-middle''. In Ref.~\cite{xu2026classical}, the combination of matrix product states and low-weight Pauli dynamics was demonstrated for a one-dimensional chaotic spin system.

In this work, we implement hybrid evolution using tree tensor networks
(TTNs)~\cite{krinitsin2025time} for the state and sparse Pauli
propagation~\cite{beguvsic2025real,rudolph2026pauli} for the observable.
For a critical-field quench of the two-dimensional transverse-field Ising
model on an $11\times11$ open lattice, initially fully polarized,
we reach $t=1.20$ by combining $t_S=t_H=0.60$, beyond the intervals
reached by either method separately in our calculations.
We compare the central-site magnetization with published
iPEPS~\cite{czarnik2019time,Dziarmaga_2022_Time}, numerical linked-cluster
expansion (NLCE)~\cite{Richter_2020_Quantum}, and X-truncated sparse
Pauli dynamics (xSPD)~\cite{beguvsic2025real} results, and examine
sensitivity to the Pauli truncation threshold.
The state and observable need not use the same representation or system
size. We demonstrate this by combining exact evolution on small periodic
clusters with Pauli propagation on a larger lattice.
Finally, we derive conditional bounds on both contributions to the
hybrid error, identifying how local state mixing can suppress operator
errors and how the distribution and coherence of Pauli contributions
can reduce sensitivity to state errors.

\paragraph{Hybrid Schr\"odinger--Heisenberg evolution.}
For a system of $L$ qubits, a Pauli string is a tensor product
$P=S_1\otimes\cdots\otimes S_L$, where $S_i\in\{I_i,X_i,Y_i,Z_i\}$. These $4^L$ strings form a complete operator basis $\mathcal{P}=\{P_n\}^{4^L}_{n=1}$.

We consider the transverse-field Ising Hamiltonian on the square lattice
\begin{equation}
    H=J\sum_{\langle i,j\rangle}X_iX_j+h\sum_i Z_i,
    \label{eq:ham}
\end{equation}
where $\langle i,j\rangle$ denotes nearest neighbours. We set $J=-1$
unless stated otherwise and use units with $\hbar=1$. For the critical-field quench, we take $h=h_c=-3.044382$. The initial state is $|\Psi_0\rangle=|0\rangle^{\otimes L}$ unless specified otherwise. The observable of interest is the magnetization $\langle Z_c(t)\rangle$ at the central site $c$.

As outlined above, we split the evolution time as $t=t_S+t_H$, evolving the state to $|\Psi(t_S)\rangle=U(t_S)|\Psi_0\rangle$ and the observable to $Z_c(t_H)=e^{iHt_H}Z_ce^{-iHt_H}$. The evolved operator can then be written as
\begin{equation}
    Z_c(t_H)=\sum_P c_P(t_H)P.
\end{equation}
The magnetization at the total time $t=t_S+t_H$ is then given by
\begin{equation}
    \langle \Psi_0| Z_c(t_S+t_H)|\Psi_0\rangle
    =\sum_P c_P(t_H)
    \langle\Psi(t_S)|P|\Psi(t_S)\rangle.
    \label{eq:hybrid}
\end{equation}
Thus, the hybrid calculation requires the Pauli strings and coefficients $c_P(t_H)$ of the evolved operator, together with their expectation values $\langle\Psi(t_S)|P|\Psi(t_S)\rangle$ in the evolved state.

We compute $Z_c(t_H)$ using Pauli propagation with the symmetric second-order Suzuki–Trotter decomposition
\begin{equation}
    U_{\delta t}
    =e^{-i h\delta t\sum_i Z_i/2}
     e^{-i J\delta t\sum_{\langle i,j\rangle}X_iX_j}
     e^{-i h\delta t\sum_i Z_i/2}.
    \label{eq:trotter}
\end{equation}
Each factor is a product of commuting Pauli rotations. 
A rotation generated by a Pauli string $P'$ acts on $P$ as
\begin{equation}
    e^{i\theta P'}Pe^{-i\theta P'}
    =\begin{cases}
        P, & [P',P]=0,\\
        \cos(2\theta)P+i\sin(2\theta)P'P, & \{P',P\}=0.
    \end{cases}
\end{equation}
Repeated application of the Trotter step~(\ref{eq:trotter}) leads to the
rapid growth in the number of Pauli strings, eventually making exact
propagation impractical. To control this growth, we discard contributions
to new Pauli strings with magnitudes below $\epsilon$, while retaining
contributions to already stored strings down to $10^{-3}\epsilon$.
At the end of each commuting layer, contributions to identical strings
are combined, and strings with total coefficients $|c_P|<\epsilon$ are
removed. This procedure is similar to those described in
Refs.~\cite{beguvsic2025real,rudolph2026pauli}, with some differences
in implementation detailed in the Supplemental Material~\cite{supplemental_material}.

To evaluate Eq.~(\ref{eq:hybrid}), we compute
$\langle\Psi(t_S)|P|\Psi(t_S)\rangle$ for every retained string.
We represent $|\Psi(t_S)\rangle$ as a TTN and write $P=\bigotimes_i P_i$,
with $P_i\in\{I,X,Y,Z\}$. We insert each local matrix $P_i$ between the
corresponding physical indices of the ket and bra tensors.
We then contract the network from the leaves towards the root.
To efficiently evaluate large numbers of Pauli strings, we cache and reuse
partial contractions for strings sharing the same Pauli factors on a subtree.
For a fixed $t_S$, each distinct string appearing throughout the operator
evolution is evaluated only once, and its expectation value is reused at
every recorded $t_H$. The coefficients $c_P(t_H)$ are independent of the
state and can likewise be reused for different $t_S$ or initial states.

\begin{figure}[t]
    \centering
    \includegraphics[width=\columnwidth]{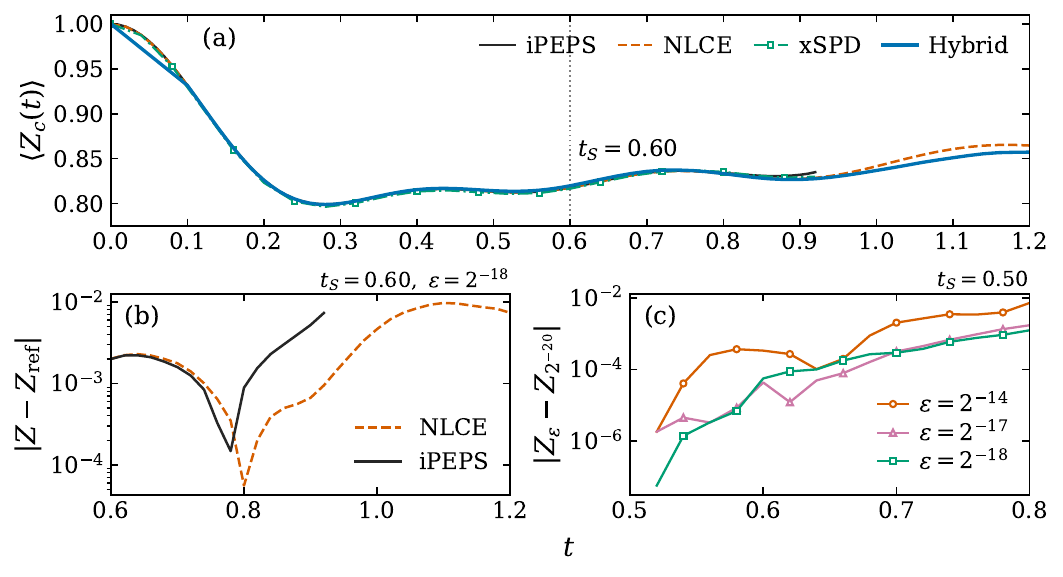}
    \caption{Hybrid dynamics of the two-dimensional transverse-field
    Ising model with $J=-1$ and $h=-3.044382$, initially polarized along
    $+Z$. (a) Central-site magnetization on an $11\times11$ open lattice,
    compared with iPEPS ($D=6$)~\cite{czarnik2019time,Dziarmaga_2022_Time},
    NLCE~\cite{Richter_2020_Quantum}, and xSPD~\cite{beguvsic2025real}.
    TTN evolution to $t_S=0.60$ is followed by hybrid evaluation with
    $\epsilon=2^{-18}$ through $t=1.20$; the dotted line marks the split.
    (b) Absolute differences of the hybrid result from NLCE and iPEPS.
    (c) Dependence on the Pauli cutoff $\epsilon$ at fixed $t_S=0.50$, relative
    to the result at $\epsilon=2^{-20}$.}
    \label{fig:magnetization}
\end{figure}

\paragraph{Two-dimensional Ising quench.}
Figure~\ref{fig:magnetization}(a) shows the central-site magnetization
after a quench to the critical field $h=-3.044382$ on an $11\times11$
open lattice, starting from the fully polarized state. We compare our
results with iPEPS~\cite{czarnik2019time,Dziarmaga_2022_Time},
NLCE~\cite{Richter_2020_Quantum}, and xSPD~\cite{beguvsic2025real}.
The blue curve is obtained from TTN evolution up to $t_S=0.60$ and then
from the hybrid evaluation using Pauli propagation with
$\epsilon=2^{-18}$ for a further $t_H=0.60$, reaching $t=1.20$
without further evolution of the state. The hybrid curve captures
the weak oscillations following the initial decrease in magnetization.
At $t=1.20$, the hybrid magnetization is approximately $0.85720$,
compared with $0.86455$ for NLCE. The difference can include numerical
errors in both calculations and finite-size effects, since the hybrid
calculation uses an $11\times11$ open lattice while NLCE describes the
thermodynamic limit. The available iPEPS data extend to
$t=0.92$~\cite{Dziarmaga_2022_Time}.

Figure~\ref{fig:magnetization}(b) compares the hybrid results with NLCE
and iPEPS over their respective available time intervals. The differences
reflect systematic errors in the methods and possible finite-size effects.

To assess cutoff sensitivity, panel (c) fixes the TTN state at $t_S=0.50$
and compares different truncation thresholds over $t_H\leq0.30$.
At $t=0.80$, the deviation from the $\epsilon=2^{-20}$ reference decreases
from $7.1\times10^{-3}$ to $1.2\times10^{-3}$ as $\epsilon$ is reduced
from $2^{-14}$ to $2^{-18}$. This demonstrates improved cutoff stability
over the tested interval, although convergence at longer times remains
to be established.

\paragraph{Mutual error protection.---}
Beyond numerical convergence, the hybrid approach raises a more fundamental
question: can the state and operator approximations suppress each other's errors? We derive conditional bounds quantifying these effects.

Let $\rho=\ket{\psi(t_S)}\bra{\psi(t_S)}$ denote the exact state and
$\tilde\rho=\ket{\phi(t_S)}\bra{\phi(t_S)}$ its TTN approximation.
For Hermitian $O$ with $\|O\|_\infty\leq1$, define
$\Delta\rho=\rho-\tilde\rho$ and $\Delta O=O(t_H)-\tilde O(t_H)$.
The hybrid error decomposes exactly as
\begin{equation}
 e(t_S,t_H)=e_S+e_H,
 \quad e_S=\operatorname{Tr}(\Delta\rho\,O(t_H)),
 \quad e_H=\operatorname{Tr}(\tilde\rho\,\Delta O).
\end{equation}
The operator probes the state error, while the state tests the operator error.

Let $\Omega$ contain the effective light cone of $X=\operatorname{supp}(O)$,
with $d_\Omega=2^{|\Omega|}$. Define the operator on $\Omega$ by
$O_\Omega=\operatorname{Tr}_{\bar\Omega}[O(t_H)]/d_{\bar\Omega}$,
so $\|O_\Omega\|_\infty\leq1$.
Write $O(t_H)=O_\Omega\otimes I_{\bar\Omega}+R_\Omega$, with
$\|R_\Omega\|_\infty\leq\epsilon_{\rm LR}$.
Lieb--Robinson locality~\cite{bravyi2006lieb} bounds this tail by
$\epsilon_{\rm LR}\leq C_{\rm LR}|X|e^{-\mu(\ell-v_{\rm LR}t_H)}$,
where $\ell$ is the distance from $X$ to the complement of $\Omega$,
and the constants depend on
the Hamiltonian and geometry. With
$\epsilon_\Omega=\|\rho_\Omega-\tilde\rho_\Omega\|_1$, one has
\begin{equation}
 |e_S|\leq|\operatorname{Tr}(\Delta\rho_\Omega O_\Omega)|+2\epsilon_{\rm LR}
 \leq\epsilon_\Omega+2\epsilon_{\rm LR}.
 \label{eq:SM-eS-split}
\end{equation}
For $\Delta O=\sum_P d_P P$, the worst-case operator bound is
$|e_H|\leq\epsilon_H=\sum_P|d_P|$.

State entanglement can suppress the operator error when the relevant reduced
states are close to maximally mixed. Define
$M=\epsilon_H^2/\sum_P|d_P|^2$ and
$\delta_A=|A|\ln2-S(\tilde\rho_A)$. Quantum Pinsker's inequality gives
$|\operatorname{Tr}(\tilde\rho P)|\leq\sqrt{2\delta_A}$ for a nonidentity
Pauli string supported on $A$. If every pair of distinct Pauli strings
$P,Q$ with nonzero error coefficients satisfies
$M\sqrt{2\delta_{\operatorname{supp}(P^\dagger Q)}}\leq\eta$, then
\begin{align}
 |e_H|^2&\leq\operatorname{Tr}(\tilde\rho\,\Delta O^\dagger\Delta O)
 \leq(1+\eta)\sum_P|d_P|^2.
\end{align}
For $\eta=O(1)$, this condition replaces the $\ell_1$ bound by an
$\ell_2$ bound, giving a suppression factor of order $M^{-1/2}$.
Related mechanisms were discussed in
Refs.~\cite{Zhao2022Random,Zhao2025Entanglement,Feng2025Trotterization,Qi2026sciadv,xu2026classical}.

To examine suppression of the state error, expand $O_\Omega=\sum_Pc_PP$,
where the Pauli strings act on $\Omega$, and set
$x_P=\operatorname{Tr}(\Delta\rho_\Omega P)$ and
$\nu^2=\sum_P|c_P|^2\leq1$. Consider an ensemble of independent,
uniformly random coefficient signs $\sigma$, keeping $|c_P|$ and $x_P$ fixed.
\textit{Lemma 1.---} If $\sum_P|c_P|^2|x_P|^2>0$, the ratio
$K(\sigma)=|\sum_Pc_Px_P|^2/\sum_P|c_P|^2|x_P|^2$ has mean 1 and variance at
most 2 in this ensemble. This random-sign result motivates, but does not
establish, analogous cancellations in physical dynamics.

Define the Pauli-weight distribution $p_P=|c_P|^2/\nu^2$,
$S_O=-\sum_Pp_P\ln p_P$, $M_O=(\sum_Pp_P^2)^{-1}$ and
$\delta S_O=2|\Omega|\ln2-S_O$. Here $S_O$ is basis-dependent Shannon
entropy, not spatial operator entanglement.
For physical coefficients, assume the coherence condition
\begin{equation}
 \left|\sum_Pc_Px_P\right|^2\leq K\sum_P|c_P|^2|x_P|^2.
 \label{eq:notes-coherence-condition}
\end{equation}
Here $K\geq1$ is a specified bound, rather than a random ensemble variable.

\textit{Theorem 1 (Mutual error protection).---} Suppose (i) the pairwise
entropy-deficit condition above holds with $\eta\geq0$; (ii)
$M_O\sqrt{2\delta S_O}\leq\zeta d_\Omega$ with $\zeta\geq0$; and (iii)
Eq.~\eqref{eq:notes-coherence-condition} holds with a specified $K\geq1$.
Then
\begin{equation}
 \begin{aligned}
 |e_H|&\leq\sqrt{1+\eta}\,\epsilon_H/\sqrt M,\\
 |e_S|&\leq\sqrt{K(1+\zeta)/2}\,\epsilon_\Omega
             \sqrt{d_\Omega/M_O}+2\epsilon_{\rm LR}.
 \end{aligned}
 \label{eq:notes-mutual-theorem}
\end{equation}
The total error is bounded by $|e_S|+|e_H|$. Zero operator error or
$\nu=0$ is handled directly without defining the corresponding ratio;
if $\sum_P|c_P|^2|x_P|^2=0$, the local state-error contribution vanishes.

These estimates formalize complementary protection mechanisms under
explicit hypotheses. For $M_O\sim d_\Omega^2$ and $K,\zeta=O(1)$, the
local state-error bound is suppressed by $d_\Omega^{-1/2}$. This regime
requires $\delta S_O=O(d_\Omega^{-2})$; extensive Pauli-weight entropy or
volume-law state entanglement alone does not establish the hypotheses.
The proofs are given in the supplement.

\paragraph{Combining exact diagonalization with Pauli propagation.---}
The Schr\"odinger-picture part of the hybrid method can also be carried
out using exact diagonalization (ED) instead of a TTN.
We evolve $4\times4$ and $5\times5$ periodic clusters up to $t_S$
and combine these states with Pauli propagation of the observable
on the $11\times11$ lattice through $t_H$.
To obtain the magnetization at $t=t_S+t_H$, we evaluate the propagated
Pauli strings in the cluster states. We keep only strings whose entire
support fits inside the cluster and discard the rest.

\begin{figure}[t]
    \centering
    \includegraphics[width=\columnwidth]{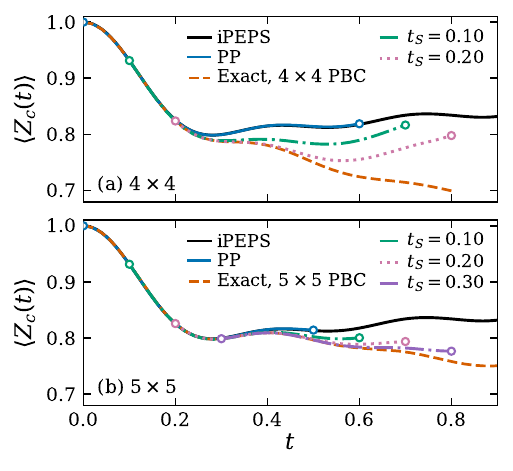}
    \caption{Hybrid dynamics using small periodic clusters for the
    quench of Fig.~\ref{fig:magnetization}.
    (a) $4\times4$ states with operator evolution through $t_H=0.60$.
    (b) $5\times5$ states with operator evolution through $t_H=0.50$.
    Each panel compares hybrids at the indicated $t_S$ with direct
    cluster evolution, iPEPS ($D=6$)~\cite{Dziarmaga_2022_Time}, and
    direct $11\times11$ OBC Pauli propagation
    ($\epsilon=2^{-20}$, $\delta t=0.02$).
    Hybrid curves are plotted at $t=t_S+t_H$; open circles mark their
    endpoints and those of direct PP.}
    \label{fig:ed-hybrid}
\end{figure}

Figure~\ref{fig:ed-hybrid} shows that the resulting hybrid curves
agree better with iPEPS at late times than exact evolution of the
clusters alone. Finite-size effects, both direct and those caused by
the omission of strings, remain, but the hybrid approach improves the
small-cluster estimate of the large-system
dynamics without increasing the cluster size, at a modest additional
computational cost. Implementation details are given in the
Supplemental Material~\cite{supplemental_material}.

\paragraph{Discussion and outlook.}
We have demonstrated that combining Schr\"odinger- and Heisenberg-picture
evolution extends the accessible time scales of quantum many-body
simulations. For the critical two-dimensional Ising quench, TTN state
evolution up to $t_S=0.60$ and Pauli propagation over $t_H=0.60$ give
magnetization dynamics up to $t=1.20$, beyond the intervals reached by
either component separately in our calculations.

The approach is not restricted to tensor-network state evolution.
Combining exact evolution of small periodic clusters with Pauli
propagation on a larger lattice improves estimates of the large-system
dynamics without increasing the cluster size. The same operator
calculation can therefore be combined with different methods for
representing and evolving the state. In particular, neural wave
functions~\cite{Sinibaldi_2026_Time-Dependent,Chen_2026_Convolutional}
may be worth exploring as an alternative for the Schr\"odinger-picture
part of the hybrid approach.

For a fixed Hamiltonian and observable, Pauli coefficients can be reused
across different initial states and preparation times. Cached
tensor-network contractions allow efficient evaluation of large
collections of Pauli strings. These evaluations are naturally
parallelizable, with a trade-off between parallel execution and reuse
of cached contractions.
State-adapted Pauli propagation instead incorporates the initial state
into operator compression through generalized mean-field
projections~\cite{perez2026generalized}. Here, state dependence enters
through contractions of a separately propagated operator.

Our theoretical analysis identifies complementary mechanisms through
which the two representations can suppress each other's approximation
errors. Under explicit conditions on local state mixing, Pauli-coefficient
distributions, and coefficient coherence, the bounds improve upon
worst-case error estimates. Their quantitative relevance to the present
simulations remains to be established.

The relative costs of state and operator evolution depend on the problem.
One representation may remain efficient well beyond the accessible
times of the other, making the optimal division between $t_S$ and $t_H$
task-dependent. Quantitative error control requires separate convergence
analysis of both approximations.
Natural extensions include fermionic dynamics combining Majorana
propagation~\cite{miller2025simulation,d2026majorana} with fermionic
tensor networks~\cite{mortier2025fermionic}, as well as adaptations to
imaginary-time evolution~\cite{GomezLurbe2026ImaginaryTime}.

\paragraph{Acknowledgments.}
Y.T. thanks the HybriLIT group at JINR for computational resources
and support for the large-scale tensor-network simulations performed on
the HybriLIT heterogeneous computing platform (LIT, JINR).
B.W. is supported by the National Natural Science Foundation of China
(92365202, 12475011, and 11921005), the National Key R\&D Program of
China (2024YFA1409002), and the Shanghai Municipal Science and Technology
Project (25LZ2601100).

\bibliographystyle{unsrt}
\bibliography{ref}

\clearpage
\onecolumngrid
\newcommand{\SupplementIncluded}{}
\ifdefined\SupplementIncluded
    \let\FinishSupplement\relax
\else
    \documentclass[aps,prl,onecolumn,superscriptaddress]{revtex4-2}
    \usepackage{graphicx}
    \graphicspath{{img/}}
    \usepackage{amsmath,amssymb}
    \usepackage{braket}
    \usepackage{booktabs}
    \usepackage{bm}
    \usepackage{xcolor}
    \usepackage{dsfont}
    \usepackage[ruled]{algorithm2e}
    \SetKwInOut{Input}{Input}
    \SetKwInOut{Output}{Output}
    \usepackage{hyperref}
    \newcommand{\FinishSupplement}{%
        \bibliographystyle{unsrt}%
        \bibliography{ref}%
        \end{document}%
    }
    \begin{document}
\fi

\section*{Supplementary Material}

\setcounter{secnumdepth}{3}
\setcounter{section}{0}
\setcounter{equation}{0}
\setcounter{figure}{0}
\setcounter{table}{0}
\renewcommand{\thesection}{S\arabic{section}}
\renewcommand{\theequation}{S\arabic{equation}}
\renewcommand{\thefigure}{S\arabic{figure}}
\renewcommand{\thetable}{S\arabic{table}}
\renewcommand{\theHsection}{S.\arabic{section}}
\renewcommand{\theHequation}{S.\arabic{equation}}
\renewcommand{\theHfigure}{S.\arabic{figure}}
\renewcommand{\theHtable}{S.\arabic{table}}

\tableofcontents

\section{Pauli propagation}
\label{sec:pauli-propagation}

For $L$ qubits, a Pauli string is a tensor product
$P=S_1\otimes\cdots\otimes S_L$, where
$S_i\in\{I_i,X_i,Y_i,Z_i\}$. The $4^L$ Hermitian strings form the
operator basis $\mathcal P$. We approximate the central-site spin
operator by retaining a subset $\mathcal P_\epsilon(t_H)$ of these strings,
\begin{equation}
    Z_c(t_H)\simeq\sum_{P\in\mathcal P_\epsilon(t_H)}c_P(t_H)P,
    \qquad Z_c(0)=Z_c.
\end{equation}
Only the retained strings and their real coefficients are stored.
Their number changes during the evolution and is controlled by the
coefficient cutoff $\epsilon$, rather than fixed in advance.
We use Pauli propagation~\cite{beguvsic2025real,rudolph2026pauli}
with the symmetric second-order Trotter step
\begin{equation}
    U_{\delta t}
    =e^{-ih\delta t\sum_i Z_i/2}
     e^{-iJ\delta t\sum_{\langle i,j\rangle}X_iX_j}
     e^{-ih\delta t\sum_i Z_i/2}.
\end{equation}
The field terms commute with one another, as do the interaction terms,
so each factor can be applied as a product of individual Pauli rotations.
The symmetric ordering has an error of order $\delta t^3$ per step,
giving second-order accuracy in $\delta t$ at fixed total time before
coefficient truncation. We record the operator after each complete step.

A rotation generated by $P'$ acts on a string $P$ as
\begin{equation}
    e^{i\theta P'}Pe^{-i\theta P'}=
    \begin{cases}
        P, & [P',P]=0,\\
        \cos(2\theta)P+i\sin(2\theta)P'P,
            & \{P',P\}=0.
    \end{cases}
    \label{eq:pp-rotation}
\end{equation}
For a half-step $Z$ layer, $P'=Z_i$ and $\theta=h\delta t/2$.
For an $XX$ layer, $P'=X_iX_j$ and $\theta=J\delta t$.
When $P$ and $P'$ anticommute, $iP'P$ is another Hermitian Pauli
string with a sign $\pm1$, which is included in its coefficient.

For each term $c_PP$, a commuting rotation leaves it unchanged.
An anticommuting rotation produces two terms, with amplitudes
$c_P\cos(2\theta)$ and $\pm c_P\sin(2\theta)$.
We apply the remaining rotations of the layer to these terms in the
same way. Different input terms can produce the same final string.
Their contributions are combined after the layer.

To limit the number of contributions, we apply a cutoff during this
expansion. A contribution $aP$ to a new string is discarded when
$|a|<\epsilon$. For an already stored string, the lower threshold is
$10^{-3}\epsilon$. This distinction matters because many small
contributions can change the coefficient of a stored string when added
together. Once a contribution is discarded, it is not propagated
through the remaining rotations. Each subsequent multiplication by
a sine or cosine would only reduce its magnitude.

After all contributions in a layer have been added, we also remove
strings with total coefficients $|c_P|<\epsilon$. Exact zeros are
discarded throughout. Thus the cutoff is applied both to individual
contributions during expansion and to the summed coefficients after
each layer. We use an absolute cutoff and impose no Pauli-weight
cutoff or cap on the number of strings.

Here an already stored string means one present when the current
group of input terms begins processing.\footnote{To limit temporary
memory, input terms are processed in groups selected by their hashes.
The stored-string set initially contains the layer input and is updated
with generated strings after each group. It stays fixed within a group.
The runs shown here use 32 groups per layer. This grouping can affect
which contributions are retained by the lower threshold.}
Algorithm~\ref{alg:pauli-propagation} summarizes one Trotter step.
Repeating this step gives the expansion at the desired $t_H$.

\begin{algorithm}[t]
\caption{One Trotter step of Pauli propagation}
\label{alg:pauli-propagation}
\Input{Current Pauli expansion, time step $\delta t$, cutoff $\epsilon$}
\Output{Updated expansion after one step}
\BlankLine
\ForEach{layer in the order $Z/2$, $XX$, $Z/2$}{
  start an empty output sum\;
  \ForEach{input term $c_PP$}{
    start with the single contribution $c_PP$\;
    \ForEach{rotation in this layer}{
      replace each surviving contribution by the one or two terms
      in Eq.~\eqref{eq:pp-rotation}\;
      discard zero contributions and those below $\epsilon$,
      or $10^{-3}\epsilon$ for already stored strings\;
    }
    add the surviving contributions to the output sum\;
  }
  combine identical strings and remove total coefficients
  $|c_P|<\epsilon$ and exact zeros\;
  use this sum as the input to the next layer\;
}
\end{algorithm}

\begin{figure}[t]
    \centering
    \includegraphics[width=\textwidth]{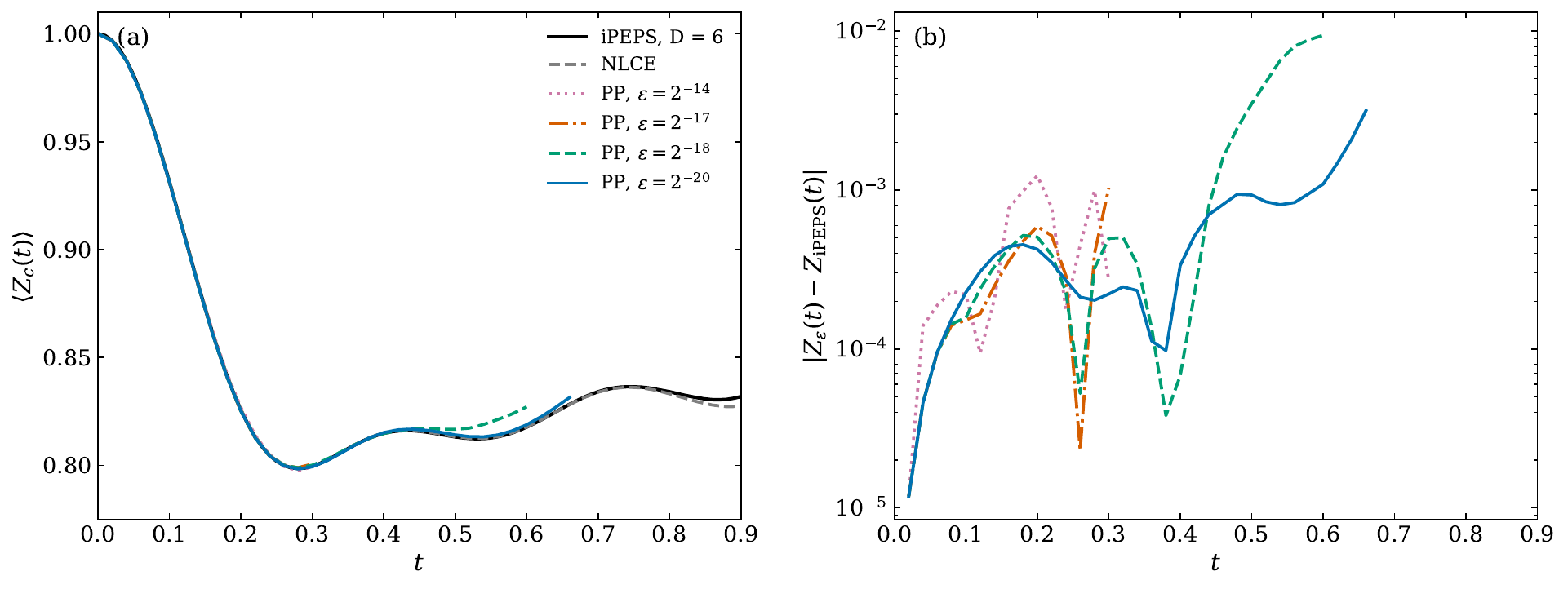}
    \caption{Direct Pauli propagation for the transverse-field Ising
    model with $J=-1$ and $h=-3.044382$, starting from the fully
    polarized state $|0\rangle^{\otimes L}$.
    (a) Central-site magnetization on an $11\times11$ open lattice at
    four coefficient cutoffs, compared with thermodynamic-limit iPEPS
    ($D=6$) and NLCE results.
    (b) Absolute differences between PP and iPEPS at matching recorded
    times. PP uses the symmetric Trotter step $\delta t=0.02$.
    Each PP curve ends at its last recorded evolution time;
    the $\epsilon=2^{-20}$ curve extends to $t=0.66$.
    The exact zero difference at $t=0$ is omitted from the logarithmic
    panel.}
    \label{fig:sup-pp}
\end{figure}

Figure~\ref{fig:sup-pp} compares direct Pauli propagation from
$|0\rangle^{\otimes L}$ with iPEPS and numerical linked-cluster
expansion (NLCE) results. Here the entire evolution is performed in
the Heisenberg picture, without a tensor-network state.
The differences from iPEPS include coefficient truncation and Trotter
errors, as well as differences between the finite open lattice and the
thermodynamic-limit reference. They do not isolate truncation error.

At each recorded operator time, the surviving strings and their
coefficients are added to a common pool. Repeated strings share the
same identifier. The pool and the coefficient lists at each time
are then used to evaluate the hybrid expectation values, as described
in Sec.~\ref{sec:combining-evolutions}.

\section{Tree tensor network representation}
\label{sec:ttn}

For $L$ qubits, a tree tensor network (TTN) represents the state
$|\psi\rangle=\sum_{s_1,\ldots,s_L}\psi(s_1,\ldots,s_L)
|s_1\cdots s_L\rangle$ by tensors connected without closed
loops~\cite{shi2006classical,tagliacozzo2009simulation}.
The $L$ leaves carry the physical indices $s_i\in\{0,1\}$.
Internal tensors have only bond indices.
Let $V_{\mathrm{int}}$ be the internal vertices,
$V=\{1,\ldots,L\}\cup V_{\mathrm{int}}$ all vertices, and
$E=\{1,\ldots,M\}$ the bonds.
Bond $e$ carries an index $\alpha_e=1,\ldots,D_e$, where $D_e$
is its bond dimension. We also label a bond by its endpoints
$e=(u,v)$.
Let $\mathcal N(v)$ denote the neighbors of $v$ and write
$\boldsymbol\alpha_v=(\alpha_{(u,v)})_{u\in\mathcal N(v)}$
for its bond indices in a fixed order.
For a leaf $i$, $p(i)$ is its only neighbor.
For $\boldsymbol s=(s_1,\ldots,s_L)$, the amplitudes are
\begin{equation}
    \begin{aligned}
    \psi(\boldsymbol s)
    ={}&\sum_{\alpha_1=1}^{D_1}\cdots\sum_{\alpha_M=1}^{D_M}
       \prod_{i=1}^{L}A^{[i]}(s_i,\alpha_{(i,p(i))})\\
    &\times\prod_{v\in V_{\mathrm{int}}}
       A^{[v]}(\boldsymbol\alpha_v).
    \end{aligned}
    \label{eq:ttn-state}
\end{equation}
Each bond index occurs in the two tensors joined by that bond and
is summed over its $D_e$ values. Cutting bond $e$ separates the
physical spins into two blocks. Their Schmidt rank is bounded by
$D_e$, which controls the entanglement that the TTN can represent.

As for a matrix product state (MPS), a TTN has a mixed-canonical form
with a chosen orthogonality
center~\cite{schollwoeck2011density,bauernfeind2020time}.
Every other tensor is orthogonalized along its path to this center,
which may be a leaf or an internal tensor.
For an internal tensor $u$ and a directed bond $u\to v$, write
the tensor orthogonalized toward $v$ as
$Q_{u\to v}(\boldsymbol\beta,\alpha)$.
Here $\alpha$ is the index on $(u,v)$, and
$\boldsymbol\beta=(\beta_{(w,u)})_{w\in\mathcal N(u)\setminus\{v\}}$
lists the indices on the other bonds at $u$.
For an internal tensor and a leaf, respectively, the isometry
conditions are
\begin{equation}
    \begin{aligned}
    \sum_{\boldsymbol\beta}
      Q_{u\to v}^*(\boldsymbol\beta,\alpha)
      Q_{u\to v}(\boldsymbol\beta,\alpha')
       &=\delta_{\alpha,\alpha'},\\
    \sum_{s_i=0}^{1}
      Q_{i\to p(i)}^*(s_i,\alpha)
      Q_{i\to p(i)}(s_i,\alpha')
       &=\delta_{\alpha,\alpha'}.
    \end{aligned}
    \label{eq:ttn-isometry}
\end{equation}
Each component of $\boldsymbol\beta$ runs over the dimension of
its bond. To obtain Eq.~\eqref{eq:ttn-isometry}, reshape
the internal tensor as $A^{[u]}_{\boldsymbol\beta,\alpha}$
and a leaf tensor as $A^{[i]}_{s_i,\alpha}$. Perform the singular
value decomposition (SVD) $A=U S V^\dagger$.
Set $Q_{u\to v}=U$ and absorb $SV^\dagger$ into the tensor at $v$.
No singular values are discarded in this change of gauge.
For example, to place the center at an internal root $r$, let $p(u)$ be the
neighbor on the path from $u$ to $r$, and orthogonalize each
$u\ne r$ toward $p(u)$, processing the most distant tensors first.
Then $Q_u\equiv Q_{u\to p(u)}$, and the remaining tensor $C_r$
is the orthogonality center.

Cut bond $(u,v)$ and consider the physical spins on the side
containing $u$. Contracting the tensors on this side, orthogonalized
toward the cut, defines the block states $|\Phi^{u\to v}_\alpha\rangle$.
For an internal $u$, their recursive definition is
\begin{equation}
    |\Phi^{u\to v}_\alpha\rangle
    =\sum_{\boldsymbol\beta}
       Q_{u\to v}(\boldsymbol\beta,\alpha)
       \bigotimes_{w\in\mathcal N(u)\setminus\{v\}}
       |\Phi^{w\to u}_{\beta_{(w,u)}}\rangle .
    \label{eq:ttn-block-states}
\end{equation}
At a physical leaf $i$, this reduces to
\[
    |\Phi^{i\to v}_\alpha\rangle
    =\sum_{s_i=0}^{1}Q_{i\to v}(s_i,\alpha)|s_i\rangle.
\]
The isometry conditions make these block states
orthonormal~\cite{bauernfeind2020time}:
\[
    \langle\Phi^{u\to v}_\alpha|\Phi^{u\to v}_{\alpha'}\rangle
       =\delta_{\alpha,\alpha'}.
\]

At an internal center $v$, the product basis is
$|\chi^{[v]}_{\boldsymbol\alpha_v}\rangle$; at a leaf center $i$,
it is $|\chi^{[i]}_{s_i,\alpha}\rangle$. Explicitly,
\begin{equation}
    \begin{aligned}
    |\chi^{[v]}_{\boldsymbol\alpha_v}\rangle
       &=\bigotimes_{u\in\mathcal N(v)}
         |\Phi^{u\to v}_{\alpha_{(u,v)}}\rangle,\\
    |\chi^{[i]}_{s_i,\alpha}\rangle
       &=|s_i\rangle|\Phi^{p(i)\to i}_{\alpha}\rangle.
    \end{aligned}
    \label{eq:ttn-local-basis}
\end{equation}
The corresponding canonical expansions are
\begin{equation}
    \begin{aligned}
    |\psi\rangle
       &=\sum_{\boldsymbol\alpha_v}
         C_v(\boldsymbol\alpha_v)
         |\chi^{[v]}_{\boldsymbol\alpha_v}\rangle,\\
    |\psi\rangle
       &=\sum_{s_i=0}^{1}\sum_\alpha
         C_i(s_i,\alpha)|\chi^{[i]}_{s_i,\alpha}\rangle .
    \end{aligned}
    \label{eq:ttn-canonical-state}
\end{equation}
The $\chi$ states are orthonormal, so
$\langle\psi|\psi\rangle=\|C_v\|_F^2$.
The center can also be placed on a bond $e=(u,v)$.
With $|\Phi^L_\alpha\rangle=|\Phi^{u\to v}_\alpha\rangle$ and
$|\Phi^R_\beta\rangle=|\Phi^{v\to u}_\beta\rangle$, the state is
\begin{equation}
    |\psi\rangle
    =\sum_{\alpha,\beta}
       C_e(\alpha,\beta)
       |\Phi^L_\alpha\rangle|\Phi^R_\beta\rangle.
    \label{eq:ttn-link-state}
\end{equation}
The matrix $C_e$ is the link tensor joining the two orthonormal
block bases. An SVD of $C_e$ yields the Schmidt decomposition across
the bond~\cite{schollwoeck2011density,bauernfeind2020time}.

\begin{figure*}[t]
    \centering
    \includegraphics[width=\textwidth]{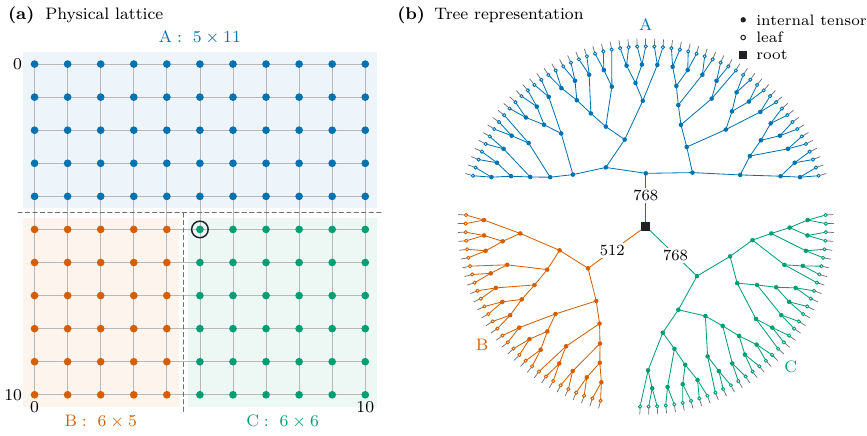}
    \caption{TTN used for the $11\times11$ open lattice.
    (a) Physical sites and nearest-neighbor interactions.
    The shaded regions identify the three root branches; the circled
    site is the central spin. Dashed lines indicate their boundaries.
    (b) The same branches in the complete tree, with all 240 tensors
    and 239 bonds shown. Filled circles denote internal tensors,
    open circles denote leaf tensors, and short outer legs carry the
    physical indices. The square is the root tensor.
    Branch labels match panel (a), and the three numbers next to the
    root are bond dimensions, as listed in
    Table~\ref{tab:ttn-parameters}. The circular placement is a drawing
    convention and does not represent physical distance.}
    \label{fig:sup-ttn}
\end{figure*}

For our $11\times11$ lattice, the root $r$ joins three branches
representing the $5\times11$,
$6\times5$, and $6\times6$ rectangles in Fig.~\ref{fig:sup-ttn}(a).
These regions are obtained by separating the top five rows from the
bottom six, then dividing the bottom rectangle into five left
columns and six right columns.
Within each region we recursively bisect rectangles along their
longer side until each leaf contains one spin.
Equal side lengths are split along the row direction; for an odd
length, the upper or left part contains one fewer row or column.
The root has three children and every other internal tensor has two.
The resulting tree has 119 internal tensors and 239 bonds.
The Hamiltonian includes all 220 horizontal and vertical
nearest-neighbor couplings in Fig.~\ref{fig:sup-ttn}(a), including
those between differently colored regions.

\section{One-site TDVP and numerical settings}
\label{sec:ttn-tdvp}

We use the one-site time-dependent variational principle
(TDVP)~\cite{haegeman2011time,bauernfeind2020time} to evolve
$|\psi(t_S)\rangle$ with fixed bond dimensions.
Updating one tensor at a time avoids the larger intermediate tensor
formed by merging neighboring tensors, reducing the peak memory
requirement~\cite{bauernfeind2020time}.
For a time-independent Hamiltonian, one-site TDVP conserves the
state norm and energy when the local exponential actions are
exact~\cite{haegeman2016unifying}. Two-site TDVP can increase the
bond dimensions; the singular-value truncation used to limit
them can change the energy. One-site evolution has no such
truncation~\cite{haegeman2016unifying,bauernfeind2020time}.
Numerical violations of norm and energy conservation arise from the
local solver and floating-point arithmetic.
We also enforce the global $\mathbb Z_2$ parity using quantum
numbers on the bond indices. This parity is preserved in both
one-site and two-site TDVP when all operations, including any
truncation, respect the parity blocks~\cite{itensor}.
Our one-site updates keep the even and odd bond dimensions fixed.

TDVP projects the Schr\"odinger equation onto the tangent space of
the TTN, the state changes obtained by varying its tensors to first
order. Writing $\mathcal P_{T_\psi}$ for the orthogonal projector onto
this space,
\begin{equation}
    \frac{d}{dt}|\psi\rangle
    =-i\mathcal P_{T_\psi}H|\psi\rangle.
    \label{eq:ttn-tdvp}
\end{equation}
The projection selects the allowed time derivative closest to
$-iH|\psi\rangle$ in the Hilbert-space norm.
Using the block states in Eq.~\eqref{eq:ttn-block-states}, define
the projector on each directed component,
\begin{equation}
    \mathcal P_{u\to v}
       =\sum_{\alpha=1}^{D_{(u,v)}}
          |\Phi^{u\to v}_\alpha\rangle
          \langle\Phi^{u\to v}_\alpha|.
    \label{eq:ttn-block-projector}
\end{equation}
The one-site and link projectors are
\begin{equation}
    \begin{aligned}
    \Pi_v&=
       \begin{cases}
       \displaystyle\bigotimes_{u\in\mathcal N(v)}
          \mathcal P_{u\to v},&v\in V_{\mathrm{int}},\\
       I_i\otimes\mathcal P_{p(i)\to i},&v=i\text{ a leaf},
       \end{cases}\\
    \Pi_e&=\mathcal P_{u\to v}\otimes\mathcal P_{v\to u},
          \qquad e=(u,v).
    \end{aligned}
    \label{eq:ttn-site-link-projectors}
\end{equation}
Here $I_i$ is the identity on the physical qubit at leaf $i$.
The tangent-space projector for a tree is
~\cite{haegeman2016unifying,bauernfeind2020time}
\begin{equation}
    \mathcal P_{T_\psi}
       =\sum_{v\in V}\Pi_v-\sum_{e\in E}\Pi_e.
    \label{eq:ttn-tangent-projector}
\end{equation}
The second sum subtracts the changes in the state counted at both
ends of each bond.
The projector-splitting integrator evolves the terms in these two
sums successively~\cite{haegeman2016unifying,bauernfeind2020time}.

For the $\Pi_v$ term, place the orthogonality center at $v$ and use
the basis in Eq.~\eqref{eq:ttn-local-basis}.
The coefficient vector $\boldsymbol C_v$ has entries
$C_v(\boldsymbol\alpha_v)$ for an internal tensor and
$C_i(s_i,\alpha)$ for a leaf.
The effective Hamiltonians in these two bases and the forward
evolution over an interval $\Delta$ are
\begin{equation}
    \begin{aligned}
    H_v^{\mathrm{eff}}(\boldsymbol\alpha,\boldsymbol\beta)
       &=\langle\chi^{[v]}_{\boldsymbol\alpha}|H|
                \chi^{[v]}_{\boldsymbol\beta}\rangle,\\
    H_i^{\mathrm{eff}}(s\alpha,t\beta)
       &=\langle\chi^{[i]}_{s,\alpha}|H|
                \chi^{[i]}_{t,\beta}\rangle,\\
    \boldsymbol C_v^{\mathrm{new}}
       &=e^{-iH_v^{\mathrm{eff}}\Delta}\boldsymbol C_v.
    \end{aligned}
    \label{eq:ttn-site-update}
\end{equation}
The matrix elements are obtained by contracting the Hamiltonian
with all tensors except $C_v$, held fixed during this local
evolution~\cite{bauernfeind2020time}.

To continue to a neighboring tensor $u$, factorize the updated
$C_v$ toward bond $e=(v,u)$. Leave the isometric factor at $v$.
The remaining factor is the link tensor $C_e$ of
Eq.~\eqref{eq:ttn-link-state}.
Using the orthonormal states on the two sides of this bond, define
\begin{equation}
    \begin{aligned}
    K_e^{\mathrm{eff}}(\alpha\beta,\alpha'\beta')
       &=\langle\Phi^L_\alpha\Phi^R_\beta|H|
                \Phi^L_{\alpha'}\Phi^R_{\beta'}\rangle,\\
    \boldsymbol C_e^{\mathrm{new}}
       &=e^{+iK_e^{\mathrm{eff}}\Delta}\boldsymbol C_e.
    \end{aligned}
    \label{eq:ttn-bond-update}
\end{equation}
Here $\boldsymbol C_e$ lists the entries of the link tensor.
The positive sign follows from the $-\Pi_e$ term in
Eq.~\eqref{eq:ttn-tangent-projector}: the link tensor is evolved
backward in time~\cite{haegeman2016unifying,bauernfeind2020time}.
Absorbing the updated link tensor into $u$ makes $u$ the next
orthogonality center.
A complete step consists of a sweep through the tree followed by
the reverse sequence, with $\Delta=\delta t/2$.
The reverse sweep also reverses the order of the local site and
link updates. This symmetric composition is second order in
$\delta t$~\cite{haegeman2016unifying,bauernfeind2020time}.
The full Hamiltonian enters the effective Hamiltonians.

Our calculations use the published one-site TDVP routines in
ITensorNetworks.jl~\cite{itensornetworks}, built on
ITensor~\cite{itensor}.
For the $11\times11$ lattice, let $n_e$ be the number of spins on
the smaller side of bond $e$. We choose $D_e$ according to $n_e$.
For example, a bond separating three spins from the other 118 has
$D_e=2^3=8$; a bond separating nine spins from the other 112 has
$D_e=384$.
Table~\ref{tab:ttn-parameters} gives all choices.
These dimensions are allocated at $t_S=0$ and held fixed throughout
the evolution.
The 217 bonds separating at most six spins retain all $2^{n_e}$
block states. The other 22 bonds have dimensions 384, 512, or 768.

\begin{table}[htbp]
\caption{Bond dimensions used from $t_S=0$.
Cutting bond $e$ divides the 121 spins into two blocks; $n_e$ is
the number of spins in the smaller block.
Only the block sizes occurring in this tree are listed.}
\label{tab:ttn-parameters}
\centering
\begin{tabular}{c@{\quad}r@{\quad}r}
\toprule
$n_e$ & $D_e$ & Number of bonds\\
\midrule
$1$       & 2   & 121\\
$2$       & 4   & 56\\
$3$       & 8   & 9\\
$4$       & 16  & 16\\
$6$       & 64  & 15\\
$9,10,15$ & 384 & 15\\
$18,25,30$& 512 & 5\\
$36,55$   & 768 & 2\\
\bottomrule
\end{tabular}
\end{table}

The initial state is the exact product state
$|0\rangle^{\otimes121}$, represented with these bond dimensions.
We conserve the even sector of $\prod_i Z_i$ using block-sparse
tensors, with equal even and odd dimensions on each bond.
The time step is $\delta t=0.01$. The local matrix exponential
actions use a Krylov solver with tolerance $10^{-10}$ and Krylov
dimension 12. All tensors have complex double-precision entries.
No singular-value truncation is performed during the one-site sweeps,
and the state is normalized after each complete step.

\section{Contraction of Pauli strings with the TTN}
\label{sec:ttn-pauli-contraction}
\label{sec:combining-evolutions}

For a Pauli string $P=S_1\otimes\cdots\otimes S_L$, we calculate
\begin{equation}
    m_P(t_S)=\frac{\langle\psi(t_S)|P|\psi(t_S)\rangle}{\mathcal N},
    \qquad \mathcal N=\langle\psi|\psi\rangle.
    \label{eq:ttn-pauli-expectation}
\end{equation}
Consider a block $B$ separated from the rest of the tree by one
bond $e$. Let $P_B=\bigotimes_{i\in B}S_i$ be the operator on its
spins, and write $|\Phi^B_\alpha\rangle$ for the block states
defined in Eq.~\eqref{eq:ttn-block-states}.
The matrix representing $P_B$ in this basis has entries
\begin{equation}
    E_B(P_B;\alpha,\beta)
       =\langle\Phi^B_\alpha|P_B|\Phi^B_\beta\rangle.
    \label{eq:ttn-block-operator}
\end{equation}
Here $\alpha,\beta=1,\ldots,D_e$, so the entire block contracts to a
$D_e\times D_e$ matrix.
Isometry cancels a branch carrying only identities. With conserved
parity, the block states also satisfy
$Z_B|\Phi^B_\alpha\rangle=p_\alpha|\Phi^B_\alpha\rangle$, where
$Z_B=\bigotimes_{i\in B}Z_i$ and $p_\alpha=\pm1$ is their parity,
specified by the bond quantum numbers~\cite{itensor}.
Orthonormality gives
\begin{equation}
    \begin{aligned}
    E_B(I;\alpha,\beta)&=\delta_{\alpha,\beta},\\
    E_B(Z_B;\alpha,\beta)&=p_\alpha\delta_{\alpha,\beta}.
    \end{aligned}
    \label{eq:ttn-identity-message}
\end{equation}
Both matrices follow without numerical tensor contractions.
The identity cancellation follows from the canonical form, as in
an MPS~\cite{schollwoeck2011density,bauernfeind2020time}.
Global parity gives $m_P=0$ for an odd number of $X$ or $Y$ factors.

These matrices are calculated from the leaves toward the open bond.
With the Pauli factors understood, write $E_{u\to v}$ for the matrix
on the $u$ side of a cut bond $(u,v)$ and $Q_i=Q_{i\to p(i)}$.
The leaf and internal-tensor contractions are
\begin{equation}
    \begin{aligned}
    E_{i\to p(i)}&=Q_i^\dagger S_i Q_i,\\
    E_{u\to v}
       &=Q_{u\to v}^\dagger
         \left(\bigotimes_{w\in\mathcal N(u)\setminus\{v\}}
            E_{w\to u}\right)Q_{u\to v}.
    \end{aligned}
    \label{eq:ttn-operator-recursion}
\end{equation}
Here each $Q$ is viewed as the matrix used in its isometry
condition, Eq.~\eqref{eq:ttn-isometry}.

Fix the cut $e$ with the center $r$ among the remaining spins
$\bar B$.
Holding the cut index at $\alpha$, contract the outside tensors,
including $C_r$, to obtain a state $|\Psi^{\bar B}_\alpha\rangle$
on $\bar B$. Then
\[
    |\psi\rangle=\sum_{\alpha=1}^{D_e}
       |\Psi^{\bar B}_\alpha\rangle|\Phi^B_\alpha\rangle.
\]
The outside states need not be normalized or orthogonal.
Contracting them with $P_{\bar B}=\bigotimes_{i\in\bar B}S_i$
gives the outside matrix
\begin{equation}
    F_{\bar B}(P_{\bar B};\alpha,\beta)
       =\langle\Psi^{\bar B}_\alpha|P_{\bar B}|
                \Psi^{\bar B}_\beta\rangle.
    \label{eq:ttn-outside-operator}
\end{equation}
Both $F_{\bar B}$ and $E_B$ have two open bond indices: $\alpha$
from the bra and $\beta$ from the ket. Joining the two sides means
multiplying entries with these same indices and summing them,
\begin{equation}
    m_P=\frac{1}{\mathcal N}\sum_{\alpha,\beta}
       F_{\bar B}(P_{\bar B};\alpha,\beta)
       E_B(P_B;\alpha,\beta).
    \label{eq:ttn-cut-contraction}
\end{equation}
For $P_{\bar B}=I$, $F_{\bar B}$ is the overlap matrix of these
outside states and retains the normalization and Schmidt weights.

To form groups, list each string's Pauli factors on $\bar B$ in
physical-site order. Two strings belong to the same group if these
lists agree at every outside site; their factors inside $B$ may
differ arbitrarily.
For example, changing $X$ to $Z$ inside $B$ preserves the group;
changing an outside factor changes it.
Large groups are processed in bounded chunks and small batches,
with several chunks active concurrently.
The common $F_{\bar B}$ is formed when needed and retained until
its chunk finishes.

Branch matrices are cached by their directed branch and Pauli
factors. Repeated requests share a cached result or the same
ongoing calculation, including requests from different chunks
and groups. Workers evaluate a new contraction when its child
matrices are available.
Cached results can be evicted and recomputed.
Within a group, contractions with one child before the other are
also reused.
For the numerical encoding $S_i=i^{x_i z_i}X_i^{x_i}Z_i^{z_i}$,
$x_i,z_i\in\{0,1\}$, we contract only $X_i^{x_i}Z_i^{z_i}$ and
restore the product of the phases at the end.

\begin{algorithm}[tbp]
\caption{Pauli-string expectation values}
\label{alg:ttn-pauli-contraction}
\Input{Canonical TTN centered outside $B$, strings $P$, fixed block $B$}
\Output{Normalized expectations $m_P$}
\BlankLine
group strings and split large groups into chunks processed in small batches\;
\ForEach{batch from the active chunks}{
  apply Eq.~\eqref{eq:ttn-identity-message} to isometric branches carrying
  identities or block parity\;
  schedule other contractions after their child matrices are available,
  Eq.~\eqref{eq:ttn-operator-recursion}\;
  \ForEach{remaining string $P$ in the batch}{
    return $m_P$ according to Eq.~\eqref{eq:ttn-cut-contraction} \;
  }
}
\end{algorithm}

On the $11\times11$ lattice, $B$ covers rows and columns $5$--$7$
in Fig.~\ref{fig:sup-ttn}(a), with $D_e=384$.
For the calculation of all 6,368,834 distinct Pauli-string expectations
at $t_S=0.60$, representative intervals gave about three strings
per second per process using
32 CPU cores: eight workers with four BLAS threads each.
This rough rate includes contraction and intermediate output after
startup and varies with contraction reuse.

Finally, these expectations enter the hybrid magnetization estimate
\begin{equation}
    \langle Z_c(t_S+t_H)\rangle
    \simeq
    \sum_{P\in\mathcal P_\epsilon(t_H)}
         c_P(t_H)m_P(t_S).
    \label{eq:ttn-hybrid-estimator}
\end{equation}
The coefficients $c_P(t_H)$ obtained by Pauli propagation are independent
of the state. For a fixed $t_S$, we evaluate
$m_P(t_S)$ once for each distinct string
in the pool. Its value is reused at every recorded $t_H$ where the
string appears, multiplied by the corresponding coefficient
$c_P(t_H)$. Thus, increasing the number of recorded operator times
does not require repeating contractions for strings already evaluated.

\section{Hybrid propagation with small-cluster states}
\label{sec:small-cluster-hybrid}

We also combine Heisenberg-picture Pauli propagation with
Schr\"odinger-picture evolution on small periodic clusters.
For the $4\times4$ cluster, we evolve the full state vector using a
scaled Taylor expansion in the even-parity sector.
At each preparation time $t_S$, we evaluate the Pauli strings of
the propagated $11\times11$ operator on this state and sum their
coefficient-weighted expectations at $t=t_S+t_H$.
For the $5\times5$ periodic cluster, Y.~Talochka computed the
Pauli-string expectation values at $t_S=0.10$, $0.20$, and $0.30$,
starting from the same fully polarized state and using the same
Hamiltonian parameters. The returned tables use a common dictionary
of 5,471,026 strings. We combine these expectations with the
$11\times11$ operator coefficients at $\epsilon=2^{-20}$ and
$\delta t=0.02$ through $t_H=0.50$.
Each string is translated into the cluster while preserving relative
positions. Strings whose entire support does not fit are omitted.

The resulting magnetization and comparison with direct cluster
evolution and iPEPS are presented in the main text. Long Pauli strings
remain sensitive to the finite cluster size, and omitting supports
that do not fit introduces an additional approximation.

\section{Proof of Lemma 1 and Theorem 1}
\label{sec:mutual-protection-proof}

\subsection{Random-sign lemma}
Since $\Delta\rho_\Omega$, $O_\Omega$ and the Pauli strings are Hermitian,
$c_P$ and $x_P$ are real. Let $c_P=\sigma_P|c_P|$, with independent
uniform signs $\sigma_P=\pm1$, and assume
$\sum_P|c_P|^2|x_P|^2>0$. Define
\begin{equation}
 w_P=\frac{|c_P|x_P}{\sqrt{\sum_Q|c_Q|^2|x_Q|^2}},
 \qquad \sum_Pw_P^2=1.
\end{equation}
Then $K(\sigma)=(\sum_Pw_P\sigma_P)^2$ and
$\mathbb E[K]=\sum_Pw_P^2=1$. Only even occurrences of each sign
survive in the fourth moment, giving
\begin{align}
 \mathbb E[K^2]&=\sum_Pw_P^4+3\sum_{P\ne Q}w_P^2w_Q^2
 =3-2\sum_Pw_P^4\leq3,\\
 \operatorname{Var}(K)&=2\left(1-\sum_Pw_P^4\right)\leq2.
\end{align}
Thus $\mathbb P(K\geq k)\leq2/(k-1)^2$ for $k>1$.
This probability is over the specified sign ensemble, with the state-error
coefficients held fixed; no independence in a physical evolution is asserted.

\subsection{Operator-evolution error}
Let $\Delta O=\sum_Pd_PP$ and
$A_{PQ}=\operatorname{supp}(P^\dagger Q)$. Distinct canonical Pauli
strings have traceless product, possibly with a phase. Quantum Pinsker's
inequality therefore gives
$|\operatorname{Tr}(\tilde\rho P^\dagger Q)|\leq\sqrt{2\delta_{A_{PQ}}}$.
Under hypothesis (i), Cauchy--Schwarz yields
\begin{align}
 |e_H|^2&\leq\operatorname{Tr}(\tilde\rho\Delta O^\dagger\Delta O)\\
 &\leq\sum_P|d_P|^2+
 \frac{\eta}{M}\sum_{P\ne Q}|d_P||d_Q|\\
 &\leq(1+\eta)\sum_P|d_P|^2
 =(1+\eta)\epsilon_H^2/M.
\end{align}
The case $\Delta O=0$ is immediate and does not require defining $M$.

\subsection{State-evolution error}
The normalized conditional expectation used in the main text is unital and
operator-norm contractive, so $\|O_\Omega\|_\infty\leq1$. The tail gives
\begin{equation}
 |e_S|\leq|\operatorname{Tr}(\Delta\rho_\Omega O_\Omega)|
             +2\epsilon_{\rm LR}.
\end{equation}
Set $p_P=|c_P|^2/\nu^2$, $u_P=1/d_\Omega^2$ and
$x_P=\operatorname{Tr}(\Delta\rho_\Omega P)$.
Since $\nu^2=\operatorname{Tr}(O_\Omega^2)/d_\Omega\leq1$,
\begin{align}
 \sum_P|c_P|^2|x_P|^2
 &\leq\frac{1}{d_\Omega^2}\sum_P|x_P|^2
       +\sum_P(p_P-u_P)|x_P|^2.
\end{align}
Parseval's identity gives
$\sum_P|x_P|^2=d_\Omega\operatorname{Tr}(\Delta\rho_\Omega^2)$.
The positive and negative eigenvalues of the traceless Hermitian
$\Delta\rho_\Omega$ each have total magnitude $\epsilon_\Omega/2$, so
\begin{equation}
 \operatorname{Tr}(\Delta\rho_\Omega^2)\leq\epsilon_\Omega^2/2.
\end{equation}
Since $M_O\leq d_\Omega^2$, the uniform term is bounded by
$\epsilon_\Omega^2d_\Omega/(2M_O)$.
For the correction, $0\leq|x_P|^2\leq\epsilon_\Omega^2$ and
$\sum_P(p_P-u_P)=0$ imply
\begin{align}
 \sum_P(p_P-u_P)|x_P|^2
 &\leq\frac{\epsilon_\Omega^2}{2}\|p-u\|_1\\
 &\leq\frac{\epsilon_\Omega^2}{2}\sqrt{2D_{\rm KL}(p\|u)}
 =\frac{\epsilon_\Omega^2}{2}\sqrt{2\delta S_O}\\
 &\leq\frac{\epsilon_\Omega^2}{2}\frac{\zeta d_\Omega}{M_O}.
\end{align}
The second line is classical Pinsker for the probability weights; it is not
a direct trace-norm bound on the observable. Combining both terms with
the coherence hypothesis gives
\begin{equation}
 |\operatorname{Tr}(\Delta\rho_\Omega O_\Omega)|^2
 \leq\frac{K(1+\zeta)}{2}\epsilon_\Omega^2\frac{d_\Omega}{M_O}.
\end{equation}
Taking the square root and adding the tail proves the state-error bound.
If $\nu=0$ or $\sum_P|c_P|^2|x_P|^2=0$, the local pairing vanishes
and the same conclusion is obtained without the undefined ratios.

\FinishSupplement

\end{document}